# FACE RECOGNITION ASSESSMENTS USED IN THE STUDY OF SUPER-RECOGNISERS

*Eilidh Noyes*[*,1] *and Alice J. O'Toole*[1]

[1] School of Behavioural and Brain Sciences, The University of Texas at Dallas, Richardson, USA

The purpose of this paper is to provide a brief overview of nine assessments of face processing skills. These tests have been used commonly in recent years to gauge the skills of perspective 'super-recognisers' with respect to the general population. In the literature, a person has been considered to be a 'super-recogniser' based on superior scores on one or more of these tests (cf., Noyes, Phillips, & O'Toole, in press). The paper provides a supplement to a recent review of super-recognisers aimed at readers who are unfamiliar with these tests. That review provides a complete summary of the super-recoginser literature to date (2017). It also provides a theory and a set of action points directed at answering the question "What is a super-recogniser?"

**Before They Were Famous Test (BTWFT)**

The Before They Were Famous Task (BTWFT) was designed to test accuracy in recognising famous faces across *age*-related changes. The task uses 56 images of famous identities, taken at a time before each celebrity was famous (usually a photograph of the celebrity that had been taken during childhood). Participants are asked to identify each of the faces. A correct identification is accepted if the participant correctly names the face, or if they provide a correct uniquely identifying fact for the celebrity. Results for this task are calculated in terms of the percentage of correct responses.

**The Cambridge Face Memory Task Long Form (CFMT Long-Form) (Russell, Duchaine, & Nakayama, 2009)**

The CFMT long-form is an extension of the CFMT (Duchaine & Nakayama, 2006), a task originally used to distinguish prosopagnosics from controls. The original CFMT task requires participants to learn six unfamiliar male faces by viewing them from different viewpoints during a learning phase. Participants are tested subsequently in three different test conditions, each progressively more difficult than the previous one. First, participants are tested on their ability to recognise the exact image that they learned of each identity. Next, they are tested on whether they can recognise the faces in new images. In the third test, they must recognise the identities in novel images to which noise has been added. Across the three tests, there are a total of 72 recognition trials.

---

[*] Corresponding Author address
   Email: eilidh.noyes@utdallas.edu

After testing just one super-recogniser on the task, it was clear to Russell et al. (2009) that the CFMT was not suitable to test super-recognisers, due to ceiling levels of performance. Therefore, a fourth section was added to create the **CFMT long-form**. This section consisted of very challenging trials—the test images of previously learned faces varied drastically from the images that were learned. The images used at training were all of good quality, and had been cropped to include only the face. Test images in the fourth section were not cropped and had added extra noise (more than in section three), or were profile images with added extra noise. To make the task even more challenging, distractor test images (image of faces that had not been learned) were repeated more often than any of the previous conditions. This increased familiarity with the distractor items made them harder to distinguish from other known (i.e., previously learned) identities. Comparison of performance (percentage accuracy) across the sections of the test confirmed that section four of CFMT long form was the most difficult of the sections (Russell et al., 2009).

**Cambridge Face Perception Test (CFPT) (Duchaine, Germine, & Nakayama, 2007)**

The CFPT was designed to assess face *perception* ability – i.e., identification of faces without a memory component. Participants are presented with a target face image (in ¾ view) and frontal images of morphs between the target and six other faces. These images were created by morphing the target face with the faces of six other individuals. The morph proportion between the individual and target faces retains differing degrees of similarity (e.g., 88%, 76%, 64%, 52% 28%) to the target face. Participants rank the six faces from one to six, in order of similarity with the target face. The correct rank order would place the face with the greatest proportion of the target image as most similar to the target face, followed by the image with the next greatest percentage of the target face and so on.

Participants are tested with both upright and inverted images (8 trial types of each of these conditions). This test design makes it possible to measure face matching through similarity ratings. The inclusion of upside-down faces also allows for testing of an inversion effect.

**The Glasgow Face Matching Task (short version) (Burton, White, & McNeill 2010)**

The Glasgow Face Matching Task (GFMT) (short version) is a standardised test of unfamiliar face matching with well-established general population norms. The task consists of the most difficult 40 trials from the Glasgow Face Matching Task (Burton, White, & McNeill, 2010). When completing the GFMT short form, participants view face image pairs and are tasked with deciding whether the images in each pair are two different images of the *same identity* or are instead images of two *different people* (see Figure 1). Results for this test are calculated in terms of the percentage of correct responses made.

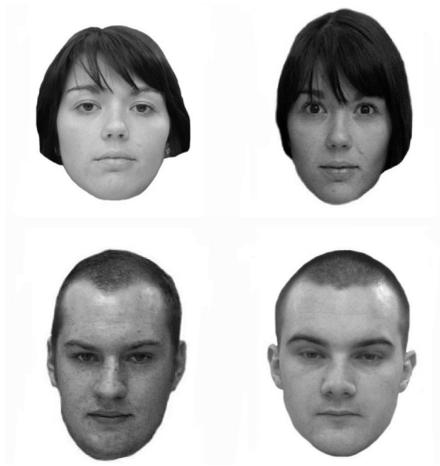

*Figure 1.* Example of a same- (top) and different-identity (bottom) trial from the GFMT. Figure image from Robertson et al. (2016).

**Models Face Matching Test (MFMT) (Dowsett & Burton, 2014)**

The Models Face Matching Test (MFMT) is similar in design to the GFMT, but consists of more challenging face stimuli. The stimuli in the models face matching test are images of models, with each image captured from a different photoshoot. These images capture large differences in appearance across multiple images of the same identity. In the MFMT participants view face image pairs and make same or different identity decisions for each image pair. As in the GFMT, this test uses percentage accuracy measures of performance.

**The Pixelated Lookalike Test (Noyes & Jenkins, 2017)**

The Pixelated Lookalike Test (PLT) is a challenging face matching task designed to assess the effect of familiarity on face matching performance for individuals of very similar appearance presented in poor quality images (Noyes & Jenkins, 2017). The task emulates that of a standard face matching test, with participants making same or different identity judgements to image pairs. Match trials in this task consist of two different images of a celebrity, whereas mismatch trials (different identity trials) consist of a celebrity image alongside an image of a professional celebrity 'lookalike' for that celebrity. This means that the people in the mismatch trials appear very similar, making this task challenging. The task is made even more difficult by presenting the images in pixelated form (see Figure 2).

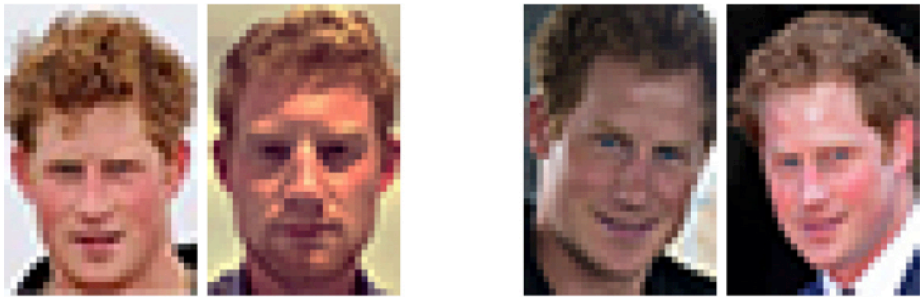

*Figure 2*. Example of the type image pairs used in the PLT. Image pair on the left shows different identities (with the imposter image on the right) and the pair on the left shows two images of the same identity. Figure from Robertson et al. (2016).

Pixelation helps to mirror the image conditions that may be encountered in poor quality surveillance footage or digital imaging devices. Following completion of the face matching task, participants are measured on their familiarity with each of the celebrities in the task. This familiarity measure allows for performance accuracy to be broken down according to personal levels of familiarity with each face. Therefore, accuracy for each individual is known for faces that are completely unfamiliar, somewhat familiar, and extremely familiar. A percentage of correct responses is thus obtained for each participant at each familiarity band.

**The Navon Test (Navon, 1977)**

The Navon test is used to provide a measure of processing bias for global compared to local image features (Navon, 1977). In the test, participants are presented with stimuli taking the appearance of an alphabetical letter, but the shape of this letter is formed by the configuration of many small letters. For example, participants may see a large 'H' on the screen but a close look at the 'H' may reveal that it is built up from many small 'T's. Participants are asked to report the large letter (in half of the trials) and the small letter (in the other half of the trials). Calculation of response error allows detection of bias to global or local features.

**Composite Face Test (Young, Hellawell & Hay, 1987)**

The composite face test is a test of holistic processing (Young, Hellawell, & Hay, 1987). Participants view composite and non-composite face images, in random order, and are asked to make identity judgments for either the top or bottom portion of the face (block dependant). A composite image consists of the top half of one face combined with the bottom half of another face, aligned into the configuration of a normal face. A non-composite image again consists of the top half of one face combined with the bottom half of another face, but misaligned horizontally. Naming accuracy is recorded and compared for the top section of the face presented in composite and non-composite form, and the bottom section of the face in composite and non-composite presentation format. Mean reaction time for accurate identification is generally lower for the non-composite image condition. This is explained by holistic processing—two identity halves presented together are 'combined' and perceived as a new identity.

**Famous Face Recognition Test (Lander et al. 2001)**

In the Famous Face Recognition Test participants are presented with 30 celebrity face images, as well as 10 unfamiliar faces (half of all images were male, the other female). Each of these images are presented one at a time and are viewed in either moving or static form (half of the images viewed in each of these conditions). The participants' task is to identify each face, by naming or providing a unique identifying fact about the face. If the face was an unknown (unfamiliar) face, the correct answer is to state that the face is unknown. After the task participant familiarity with the familiar faces is verified by providing them with a list of names of the faces that they had seen and asking whether they should have recognised the face of that celebrity. Percentage accuracy of response is calculated.